\RequirePackage{lineno}
\documentclass[10pt]{IEEEtran} 

\usepackage{epsfig,amsmath,amsfonts,cite}
\usepackage{threeparttable}
\usepackage{lipsum}
\usepackage{color}
\usepackage{graphicx}
\usepackage{lipsum}
\usepackage{algorithm,algorithmic}
\usepackage{epsfig,amsmath,amsfonts,cite}
\usepackage{amsthm}
\usepackage{booktabs}
\newcommand{\thickhline}{\noalign{\hrule height 1.0pt}}
\DeclareMathAlphabet\mathbfcal{OMS}{cmsy}{b}{n}

\usepackage{multirow}
\usepackage{multicol}
\ifCLASSINFOpdf
 
\else
\fi

\begin{document}
\title{Probabilistic Power Flow Computation via Low-Rank and Sparse Tensor Recovery}
\author{Zheng Zhang, Hung Dinh Nguyen, Konstantin Turitsyn and Luca Daniel
\thanks{Z. Zhang was with Massachusetts Institute of Technology (MIT), Cambridge, MA 02139, USA. He is now with Argonne National Laboratory, Lemont, IL 60439. E-mail: z\_zhang@mit.edu.}
\thanks{H. D. Nguyen, K. Turitsyn and L. Daniel are with MIT, Cambridge, MA 02139, USA. E-mails: \{hunghtd, turitsyn, luca\}@mit.edu.}
}%

\markboth{IEEE Transactions on Power Systems,~Vol.~, No.~, xx~20xx}%
 {Shell \MakeLowercase{\textit{et al.}}: Stochastic Power Flow}%

\maketitle

\begin{abstract}

This paper presents a tensor-recovery method to solve probabilistic power flow problems. Our approach generates a high-dimensional and sparse generalized polynomial-chaos expansion that provides useful statistical information. The result can also speed up other essential routines in power systems (e.g., stochastic planning, operations and controls).

Instead of simulating a power flow equation at all quadrature points, our approach only simulates an extremely small subset of samples. We suggest a model to exploit the underlying low-rank and sparse structure of high-dimensional simulation data arrays, making our technique applicable to power systems with many random parameters. We also present a numerical method to solve the resulting nonlinear optimization problem.

Our algorithm is implemented in MATLAB and is verified by several benchmarks in MATPOWER $5.1$. Accurate results are obtained for power systems with up to $50$ independent random parameters, with a speedup factor up to $9\times 10^{20}$.

\end{abstract}


\section{Introduction}

\IEEEPARstart{R}{ealistic} power systems are affected by various uncertainties, such as the randomness of generations and loads, insufficient knowledge about network parameters, and noisy measurement~\cite{Conejo2010Decision, borkowska1974probabilistic, haesen2007stochastic, saunders2014point, morales2007point, morales2010, Verbic:2006, Su:2005, Emil:2011, Allan:1981,Allan:1977,Miranda:1990,GLin:2014,Hockenberry:2004}. Uncertainties may increase in future power systems, since many renewables highly depend on the uncertain weather conditions~\cite{Emil:2011, morales2010}. These uncertainties must be considered in simulation, such that subsequent tasks can be completed in an efficient and robust way.

This work investigates the probabilistic power flow problem~\cite{borkowska1974probabilistic}, which quantifies the uncertainties of bus voltages and line flows under uncertain loads, generations or network parameters. Currently, this problem is routinely solved in a number of decision-making procedures. Examples include transmission expansion and planning under long-term uncertainties in renewables penetration and regulation policies \cite{milligan2012stochastic, 221258}. In operations, the operators assess the security of the system and calculate Available Transfer Capability using random scenario sampling \cite{stahlhut2005uncertain} where the ability to average the steady state solution over a large number of random scenarios is essential for secure power operations. 

Probabilistic power flow problems have been solved by Monte Carlo and many analytical methods (including multi-linearization~\cite{Allan:1981}, the comulant method~\cite{Allan:1977}, fuzzy load flow analysis~\cite{Miranda:1990}, and so forth). Recently, point estimation has become a popular technique for probabilistic power flow analysis~\cite{saunders2014point, morales2007point, morales2010,Verbic:2006,Su:2005}. This method assumes the solution being a summation of some univariate functions, then it computes the moments using a set of one-dimensional quadrature points. 

Stochastic spectral methods~\cite{xiu2009} have emerged as a promising technique for the uncertainty analysis of many engineering problems including power systems~\cite{GLin:2014,Hockenberry:2004, zzhang:tcad2013, zzhang:huq_tt}. They approximate the stochastic solution by a generalized polynomial-chaos expansion~\cite{gPC2002}. This representation can provide various statistical information (e.g., moments and probability density function); it  can also accelerate many stochastic problems in power systems (e.g., stochastic unit commitment~\cite{Emil:2011} and parameter inference~\cite{zzhang:2015_map}), whereas previous approaches generally cannot. However, stochastic spectral methods may require lots of basis functions and simulation samples for problems with many uncertainties. In the uncertainty quantification community, some techniques based on compressed sensing~\cite{yxiu:2013,JPeng:2014}, proper generalized decomposition~\cite{Nouy:2010,Chinesta:2011} and  tensor-train decomposition~\cite{tensor:gelerkin,Marzouk:2014,zzhang:huq_tt} have been developed for high-dimensional problems. 

This paper develops an alternative stochastic spectral method to solve probabilistic power flow problems with possibly high-dimensional random parameters. Our main contributions are summarized as the following: i) We use tensors~\cite{tensor:suvey} (i.e., high-dimensional data arrays) to represent the huge set of data samples required in stochastic simulation. With a tensor format, we propose a low-rank and sparse tensor recovery scheme to generate a high-dimensional and sparse approximation while using an extremely small subset of quadrature samples. ii) We present the detailed numerical implementation of  the tensor recovery method. Our algorithm relies on alternating minimization and the alternating direction method of multipliers (ADMM)~\cite{Boyd:ADMM2010}. Although only locally optimal solutions are guaranteed, the developed solver performs well for many practical cases. We demonstrate the performance of the proposed technique with numerical simulations on $3$ benchmarks in MATPOWER $5.1$~\cite{matpower:2011}.

\section{Problem formulation} \label{sec:formulation}
\subsection{Probabilistic Power Flow Problem}
A steady-state power system with uncertainties can be described with parameterized power flow equations:
\begin{equation}
 \label{eq:pf}
\begin{array}{l}
P_i (\boldsymbol{\xi})= \sum \limits_{k=1}^n {V_i(\boldsymbol{\xi})V_k(\boldsymbol{\xi})\left(G_{ik}\cos{\theta_{ik}(\boldsymbol{\xi})}+B_{ik}\sin{\theta_{ik}(\boldsymbol{\xi})}\right)}\\
Q_i (\boldsymbol{\xi})= \sum \limits_{k=1}^n{V_i(\boldsymbol{\xi})V_k(\boldsymbol{\xi})\left(G_{ik}\sin{\theta_{ik}(\boldsymbol{\xi})}-B_{ik}\cos{\theta_{ik}(\boldsymbol{\xi})}\right)}
\end{array}
\end{equation} 
where $P_i$, $Q_i$, $V_i$, $\theta_{i}$ are the active and reactive power, voltage magnitude and angle at load bus $i$, respectively; $G_{ik}$ and $B_{ik}$ are conductances and susceptances; $\theta_{ik}=\theta_i-\theta_k$ is the voltage angle difference between buses $i$ and $k$.

We employ random parameters $\boldsymbol{\xi}$$=$$[\xi_1, \cdots, \xi_d]\in \mathbb{R}^d$ to describe the uncertainties of load power consumptions that further influence bus voltages and angles. After computing $V_i$'s and $\theta_i$'s, an out of interest $y$ (e.g., the line flows) can be easily extracted. Obviously $y$ also depends on $\boldsymbol{\xi}$ and thus can be written as $y=g(\boldsymbol{\xi})$. We assume that a deterministic  solver is available to solve \eqref{eq:pf} given a sample of $\boldsymbol{\xi}$. For simplicity, we assume that all elements of $\boldsymbol{\xi}$ are mutually independent, then their joint probability density function is $\rho(\boldsymbol{\xi})=\prod \limits_{k=1}^d {\rho_k(\xi_k)} $, where $\rho_k(\xi_k)$ is the marginal probability density function of $\xi_k$. Moreover, the slack bus is assigned to compensate for the variations of loads and losses. 

\subsection{Stochastic Collocation Method} 
If the power flow problem is solvable, and $y$$=$$g(\boldsymbol{\xi}) $ smoothly depends on $\boldsymbol{\xi}$, then we can approximate $y$ by a truncated generalized polynomial-chaos expansion~\cite{gPC2002}
\small
\begin{equation}
\label{surrogate_i}
y=g\left (\boldsymbol{\xi}\right )\approx \sum\limits_{|\boldsymbol{\alpha} | \leq p} {c_{\boldsymbol{\alpha}} \Psi _{\boldsymbol{\alpha}}  (\boldsymbol{\xi})},\; {\rm with}\; \Psi _{\boldsymbol{\alpha}} (\boldsymbol{\xi})=\prod\limits_{k=1}^d {\phi_{k,\alpha_k}(\xi_k)}. 
\end{equation} \normalsize
The multivariate polynomial basis $\Psi _{\boldsymbol{\alpha}}(\boldsymbol{\xi})$ is indexed by $\boldsymbol{\alpha}$$=$$[\alpha_1, \cdots, \alpha_d] $$\in $$\mathbb{N}^d$, with the total polynomial degree $|\boldsymbol{\alpha}|$$=$$\sum\limits_{k=1}^d{|\alpha_k|}\leq p$. The total number of basis functions is 
\small
\begin{equation}
\label{Kvalue}
K = \frac{{(p + d)!}}{{p!d!}}.
\end{equation} \normalsize

As shown in Appendix~\ref{subsec:uni_gPC}, the degree-$\alpha_k$ univariate polynomials $\{\phi_{k,\alpha_k}(\xi_k)\}_{\alpha_k=0}^p$ are orthonormal to each other. Therefore, the multivariate basis functions are  also orthonormal, and $c_{\boldsymbol{\alpha}}$ can be computed with projection
\begin{equation}
\label{c_project}
c_{\boldsymbol{\alpha}}=\int\limits_{\mathbb{R}^d} {\Psi_{\boldsymbol{\alpha}} ( \boldsymbol{\xi} ) g ( \boldsymbol{\xi} )\rho(\boldsymbol{\xi}) d\boldsymbol{\xi} }.
\end{equation}
This integral can be evaluated by a proper quadrature rule which requires computing $g(\boldsymbol{\xi})$ at  a set of samples. 

\subsection{Integration Rules and Curse of Dimensionality}
\label{subsec:tensor_gq}
Among different quadrature rules~\cite{Gerstner:1998, MCintro,QMC:1995}, this work considers computing $c_{\boldsymbol{\alpha}}$ by a tensor-rule Gauss quadrature method.  First, use Gauss quadrature~\cite{Golub:1969} (in Appendix~\ref{app:gauss_quad}) to decide $m$ quadrature samples and weights  $\left\{\xi_k^{i_k}, w_k^{i_k} \right\}_{i_k=1}^{m}$ for $\xi_k$.  Next, we compute $c_{\boldsymbol{\alpha}}$ by a tensor rule 
\begin{equation}
\label{c_tp_int}
c_{\boldsymbol{\alpha}}\approx \sum\limits_{ i_1 =1}^{m} \cdots \sum\limits_{i_d =1}^{m}{ g (\xi_1^{i_1}, \cdots, \xi_d^{i_d})  \prod\limits_{k = 1}^d {\phi_{k,\alpha_k}(\xi_k^{i_k})w_k^{i _k}}}.
\end{equation}

This method requires simulating the power flow equation $m^d$ times, and obviously it only works well for low-dimensional problems (e.g., when $d$ is below $5$ or $6$). Sparse grid has been applied to simulate power systems~\cite{GLin:2014}, which can compute (\ref{c_project}) with about $2^pK$ samples for high-dimensional cases~\cite{zzhang:tcad2013}. In this paper, we aim to use only $<K$ samples from a tensor rule to compute (\ref{c_project}). 
 




\section{A Tensor-Recovery Approach}
\label{sec:tensorrecovery}
This section presents our tensor-recovery method to solve high-dimensional probabilistic power flow problems.

\subsection{Tensor Representations of~\eqref{c_tp_int}}
\label{subsec:tensor-repres}
\begin{figure}[t]
	\centering
		\includegraphics[width=2.8in]{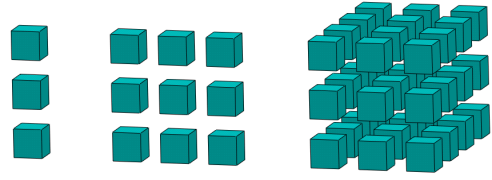} 
\caption{Demonstration of vectors (left), matrices (middle) and tensors (right).}
	\label{fig:tensor}
\end{figure}

As a generalization of vectors and matrices, a tensor $\mathbfcal{A}\in \mathbb{R}^{m_1\times \cdots \times m_d} $ represents a high-dimensional data array~\cite{tensor:suvey}. The number of dimensions, $d$, is called the mode of a tensor; $m_k$ is the size of the $k$-th dimension. Given index $\mathbf{i}=(i_1, \cdots, i_d)$ (with integer $i_k \in [1, m_k]$), we can specify one element $\mathbfcal{A}(\mathbf{i})$. Fig.~\ref{fig:tensor} shows a $1$-mode tensor (i.e., vector), a $2$-mode tensor (i.e., matrix) and a $3$-mode tensor. 

First, we define a $d$-mode tensor $\mathbfcal{G}\in \mathbb{R}^{m\times \cdots \times m} $
\begin{equation}
\mathbfcal{G}(\mathbf{i})= g (\xi_1^{i_1}, \cdots, \xi_d^{i_d}).
\end{equation}
Next, for every $\xi_k$ and its degree-$\alpha_k$ polynomial $\phi_{k,\alpha_k}(\xi_k)$, we define a vector $\mathbf{w}_{\alpha_k}^{(k)}\in \mathbb{R}^{m}$ with its $i_k$-th element being
\begin{equation}
\mathbf{w}_{\alpha_k}^{(k)}(i_k)=\phi_{k,\alpha_k}(\xi_k^{i_k})w_k^{i_k}. 
\end{equation}
For every index vector $\boldsymbol{\alpha}$, we further construct a $d$-mode \textbf{rank-1 tensor} $\mathbfcal{W}_{\boldsymbol{\alpha}}\in \mathbb{R}^{m\times  \cdots \times m} $:
\begin{equation}
\label{outer}
\mathbfcal{W}_{\boldsymbol{\alpha}}=\mathbf{w}_{\alpha_1} ^{(1)}\circ \cdots  \circ \mathbf{w}_{\alpha_d}^{(d)}\; \Leftrightarrow\; \mathbfcal{W}_{\boldsymbol{\alpha}}(\mathbf{i})=\prod\limits_{k=1}^d {\mathbf{w}_{\alpha_k} ^{(k)}(i_k)}.
\end{equation}
Here $\circ$ denotes an outer product. As a result, the right-hand side of (\ref{c_tp_int}) is the \textbf{inner product} of $\mathbfcal{G}$ and $\mathbfcal{W}_{\boldsymbol{\alpha}}$:
\begin{equation}
\label{c_tp_inner}
\boxed{c_{\boldsymbol{\alpha}}\approx \left\langle \mathbfcal{G}, \mathbfcal{W}_{\boldsymbol{\alpha}} \right\rangle =\sum\limits_{ i_1 =1}^{m} \cdots \sum\limits_{i_d =1}^{m}{ \mathbfcal{G}(\mathbf{i})  \mathbfcal{W}_{\boldsymbol{\alpha}}(   \mathbf{i})}.}
\end{equation}

In summary, in order to obtain the generalized polynomial-chaos approximation (\ref{surrogate_i}) we need to compute: 1) tensor $\mathbfcal{G}$ ; 2) tensor $\mathbfcal{W}_{\boldsymbol{\alpha}}$ for each $\boldsymbol{\alpha}$ satisfying $|\boldsymbol{\alpha}| \leq p$. Since each $\mathbfcal{W}_{\boldsymbol{\alpha}}$ is the outer product of $d$ vectors and many of them can be reused, computing $\mathbfcal{W}_{\boldsymbol{\alpha}}$'s is trivial. However, directly computing $\mathbfcal{G}$ is almost impossible, since the power flow equation must be simulated $m^d$ times. 

\subsection{Low-Rank and Sparse Tensor-Recovery}
\label{subsec:tensor-recovery}

Instead of computing $\mathbfcal{G}$ directly, we {\it approximate }$\mathbfcal{G}$ by tensor recovery. The key idea is described below.

\subsubsection{Sub-Sampling} We randomly compute a small portion of elements in $\mathbfcal{G}$, then seek for a tensor $\hat{\mathbfcal{G}}$ to approximate $\mathbfcal{G}$. Let ${\cal I}=\left \{ \mathbf{i}| 1\leq i_k \leq m\right \}$ include the indices for all elements in $\mathbfcal{G}$. The size of ${\cal I}$, $|{\cal I}|$, is $m^d$. We choose a subset $\Omega \subset {\cal I}$ (with $|\Omega |\ll |{\cal I}|$) that includes a small number of indices randomly selected from ${\cal I}$, and compute $\mathbfcal{G}(\mathbf{i})=g(\xi_1^{i_1},\cdots, \xi_d^{i_d})$  for any $\mathbf{i}\in \Omega$. Then, we look for a tensor $\hat{\mathbfcal{G}}$ such that it matches $\mathbfcal{G}$ at all elements specified by $\Omega$, i.e., 
\begin{equation}
\label{tensor_point_match}
\| \mathbb{P}_{\Omega}\left(\hat{\mathbfcal{G}} -{\mathbfcal{G}}\right)\|_F^2 =0.
\end{equation}
Here $\mathbb{P}_{\Omega}$ is a linear operator for tensors:
\begin{equation}
\label{tensor_project}
\mathbfcal{B}=\mathbb{P}_{\Omega}\left({\mathbfcal{A}}\right) \; \Leftrightarrow\;  \mathbfcal{B}(\mathbf{i}) = \left\{ \begin{array}{l}
 \mathbfcal{A}(\mathbf{i}) ,\;{\rm{if}}\;\mathbf{i} \in  {\Omega} \\ 
 0 ,\;{\rm{otherwise}}.
 \end{array} \right.
 \end{equation}
The Frobenius-norm of a general tensor is defined as
\begin{equation}
\label{tensor_F_norm}
\| \mathbfcal{A} \|_F=\sqrt{\left\langle \mathbfcal{A}, \mathbfcal{A} \right\rangle }.
\end{equation}
An infinite number of tensors exist that satisfies the requirement \eqref{tensor_point_match} but significantly differs from $\mathbfcal{G}$. Therefore, some constraints can be added to regularize this problem.

\subsubsection{Constraint $1$-- Sparsity} Let vector $\mathbf{c} =[\cdots, c_{\boldsymbol{\alpha}}, \cdots]\in \mathbb{R}^K$ includes all coefficients in the generalized polynomial-chaos approximation. In high-dimensional cases, $\mathbf{c}$ is generally very sparse -- most of its elements are close to zero. Using $\it l_1$-norm as a measure of sparsity~\cite{Donoho:2006}, we have
\begin{equation}
\label{tensor_sparse}
|\mathbf{c}|=\sum\limits_{{\boldsymbol{\alpha}}\leq p} {|c_{\boldsymbol{\alpha}}|}\approx \sum\limits_{{\boldsymbol{\alpha}}\leq p} {\left|  \left\langle \hat{\mathbfcal{G}}, \mathbfcal{W}_{\boldsymbol{\alpha}} \right\rangle \right |} \; {\rm is}\; {\rm small}
\end{equation}

\subsubsection{Constraint $2$--Low Tensor Rank} In many cases, $\mathbfcal{G}$ has a low tensor rank and can be well approximated by the summation of a few rank-$1$ tensors. Therefore, we assume that the solution $\hat{\mathbfcal{G}}$ has a \textbf{rank-$r$ decomposition}:
\begin{equation}
\label{tensor_lr}
\hat{\mathbfcal{G}} =\mathbb{T}  \left( \mathbf{U}^{(1)}, \cdots, \mathbf{U}^{(d)} \right) :=\sum\limits_{j=1}^{r} {\mathbf{u}_j^{(1)} \circ \cdots \circ \mathbf{u}_j^{(d)} }
\end{equation}
where $\mathbf{u}_j^{(k)}$$\in $$\mathbb{R}^{m \times 1}$ is the $j$-th column of matrix $\mathbf{U}^{(k)}\in \mathbb{R}^{m \times r}$. Therefore, we may use $d$ matrices to represent the whole tensor instead of computing and storing all elements of $ \hat{\mathbfcal{G}}$.

\subsubsection{Final Tensor-Recovery Model} We describe the low-rank and sparse tensor-recovery model as follows:

{ \it Given $\mathbfcal{G}(\mathbf{i})$ for every $\mathbf{i}\in \Omega$,  solve 
\begin{equation}
\label{eq:t_recovery}
\boxed{
\begin{array}{l}
 \mathop {\min }\limits_{\left\{ \mathbf{U}^{(k)}  \in \mathbb{R}^{m  \times r} \right\}_{k = 1}^d }  f\left( \mathbf{U}^{(1)} , \cdots ,\mathbf{U}^{(d)} \right) \\
\;\;\;\;\;\;\;\;\;\;\;\;\; =\frac{1}{2} \left\| {\mathbb{P}_{\Omega} \left( {\mathbb{T}( {\mathbf{U}^{(1)} , \cdots ,\mathbf{U}^{(d)} } )}- \mathbfcal{G}\right)} \right\|_F^2  \\
\;\;\;\;\;\;\;\;\;\;\;\;\;\;\;\;\;+ \lambda \sum\limits_{\left| \boldsymbol{\alpha}  \right| \le p} {\left| {\langle {\mathbb{T}( \mathbf{U}^{(1)} , \cdots ,\mathbf{U}^{(d)}  ),\mathbfcal{W}_{\boldsymbol{\alpha}  } }\rangle } \right|} .
  \end{array}}
  \end{equation} 
Here $\lambda>0$ is a regularization parameter.}

\subsection{Summary of Main Steps}
We summarize the main steps of our approach as below.
\subsubsection{Simulation Step} Randomly generate a small subset $\Omega \subset {\cal I}$ such that $|\Omega|<K \ll m^d$. For every index $\mathbf{i} =[i_1, \cdots , i_d]\in \Omega$, simulate the power flow equation \eqref{eq:pf} once to obtain a deterministic value $\mathbfcal{G}(\mathbf{i})=g(\xi_1^{i_1}, \cdots, \xi_d^{i_d})$.

\subsubsection{Optimization Step} Solve \eqref{eq:t_recovery} to obtain matrices $\mathbf{U}^{(1)}, \cdots, \mathbf{U}^{(d)}$ that represent tensor $\hat{\mathbfcal{G}}$ in \eqref{tensor_lr}.

\subsubsection{Model Generation} Replace $\mathbfcal{G}$ by $\hat{\mathbfcal{G}}$,  and calculate $c_{\boldsymbol{\alpha}}$'s according to (\ref{c_tp_inner}). With low-rank tensor factors, the computation can be simplified to
\begin{equation}
\label{eq:tensor2gPC}
c_{\boldsymbol{\alpha}} \approx \left\langle \hat{\mathbfcal{G}}, \mathbfcal{W}_{\boldsymbol{\alpha}} \right\rangle =\sum\limits_{j=1}^r {\left[ \prod\limits_{k=1}^d{\left(\left(\mathbf{u}_j^{(k)} \right)^T \mathbf{w}_{\alpha_k}^{(k)}  \right)}\right]},
\end{equation}
which involves only cheap vector inner products. 

Since we can approximate $\mathbfcal{G}$ by using only a small number of simulation samples, our method can be applied to many high-dimensional problems.

\section{Optimization Solver}
\label{sec:opt_solver}
This section describes how to solve (\ref{eq:t_recovery}). 
\begin{algorithm}[t]
\caption{Alternating Minimization for Solving (\ref{eq:t_recovery}).}
\label{alg:alm}
\begin{algorithmic}[1]
\STATE {Initialize:  $\mathbf{U}^{(k),0}\in \mathbb{R}^{m\times r}$ for $k=1,\cdots d$;}
\STATE {\textbf{for} $l=0,1,\cdots$}
  \STATE {\hspace{10pt} {\textbf{for} $k=1,\;\cdots, d$  \textbf{do}}}
   \STATE {\hspace{20pt} solve (\ref{eq:alm_k}) by Alg. \ref{alg:admm} to obtain $\mathbf{U}^{(k),l+1}$ };
   \STATE {\hspace{10pt} {\textbf{end for} } }
   \STATE {\hspace{10pt} \textbf{break} if converged;}
  \STATE {\textbf{end for} } 
\STATE {\textbf{return} $\mathbf{U}^{(k)}=\mathbf{U}^{(k),l+1}$ for $k=1,\cdots, d$. }
\end{algorithmic}
\end{algorithm}

\subsection{Outer Loop: Alternating Minimization}
\subsubsection{Algorithm Flow} Starting from an initial guess $\{\mathbf{U}^{(k),0}\}_{k=1}^d$, we perform the following iterations: at iteration $l+1$ we use $\{\mathbf{U}^{(k),l}\}_{k=1}^d$ as an initial guess and obtain updated tensor factors $\{\mathbf{U}^{(k),l+1}\}_{k=1}^d$ by alternating minimization. Each iteration consists of $d$ steps, and at the $k$-th step, ${\mathbf{U}^{(k),l+1}}$ is obtained by solving
\small
\begin{equation}
\label{eq:alm_k}
\boxed{{\mathbf{U}^{(k),l+1}}=\arg \min_{\mathbf{X}} { f\left( \cdots, \mathbf{U}^{(k-1),l+1}, \mathbf{X}, \mathbf{U}^{(k+1),l},\cdots \right)}.} 
\end{equation} \normalsize
Since all factors expect $\textbf{U}^{(k)}$ are fixed, (\ref{eq:alm_k}) becomes a convex optimization problem, and its global minimum can be computed by the solver in Section~\ref{subsec:admm}. The alternating minimization method ensures that the cost function decreases monotonically to a local minimal. The pseudo codes are summarized in Alg.~\ref{alg:alm}, which terminates when the convergence criteria in Appendix~\ref{app:errDef} is satisfied.

\subsubsection{Prediction Error} An interesting question is: how accurate is $\hat{\mathbfcal{G}}$ compared with the exact tensor $\mathbfcal{G}$? Our tensor recovery formulation enforces consistency between $\hat{\mathbfcal{G}}$ and $\mathbfcal{G}$ at the indices specified by $\Omega$. We hope that $\hat{\mathbfcal{G}}$ also has a good predictive behavior -- $\hat{\mathbfcal{G}}(\mathbf{i})$ is also close to ${\mathbfcal{G}}(\mathbf{i})$ for $\mathbf{i}\notin \Omega$. In order to measure the predictive property of our results, we define a heuristic prediction error
\begin{equation}
\epsilon_{\rm pr}=\sqrt{\frac{\sum \limits_{\mathbf{i}\in \Omega'}{\left(\hat{\mathbfcal{G}}(\mathbf{i}) - {\mathbfcal{G}}(\mathbf{i}) \right)^2 w_{\mathbf{i}}}}  {\sum \limits_{\mathbf{i}\in \Omega'}{\left( {\mathbfcal{G}}(\mathbf{i}) \right)^2 w_{\mathbf{i}}}}}, \; {\rm with}\; w_{\mathbf{i}}=\prod \limits_{k=1}^d {w_k^{i_k}}.
\end{equation}
Here $\Omega' \in {\cal I}$ is a small-size index set such that $\Omega' \cap \Omega=\emptyset$. Obviously, $\hat{\mathbfcal{G}}$ has good predictive behavior if $\epsilon_{\rm pr}$ is small. Estimating $\epsilon_{\rm pr}$ requires simulating the power flow equation at some extra quadrature samples. However, a small-size $\Omega' $ can provide a good heuristic estimation.

\subsection{Inner Loop: Numerical Solver for (\ref{eq:alm_k})}
\label{subsec:admm}

Following the procedures in Appendix~\ref{app:glasso}, we rewrite Problem (\ref{eq:alm_k}) as the generalized LASSO problem:
\begin{equation}
\label{eq:glasso}
\boxed{
\boldsymbol{vec}\left(\mathbf{U}^{(k),l+1} \right) =\arg \min_{\mathbf{x}} { \frac{1}{2} \left\| \mathbf{A}\mathbf{x}-\mathbf{b} \right\|_2^2 +\lambda | \mathbf{Fx}| }}
\end{equation}
where $\mathbf{A}\in \mathbb{R}^{|\Omega| \times mr}$, $\mathbf{F}\in \mathbb{R}^{K \times mr}$ and $\mathbf{b}\in \mathbb{R}^{ |\Omega| \times 1}$, and $\mathbf{x}=\boldsymbol{vec}(\mathbf{X})\in \mathbb{R}^{mr \times 1}$ is the vectorization of $\mathbf{X}$ (i.e., $\mathbf{x}(jm-m+i)= \mathbf{X} (i,j)$ for any integer $1\leq i\leq m$ and $1\leq j \leq r$). Note that $|\Omega|$ is the number of simulations samples in tensor recovery, and $K$ is the total number of basis functions. 

We solve (\ref{eq:glasso}) by the alternating direction method of multipliers (ADMM)~\cite{Boyd:ADMM2010}. Problem (\ref{eq:glasso}) can be rewritten as
\begin{equation}
\min_{\mathbf{x,z}} { \frac{1}{2} \left\| \mathbf{A}\mathbf{x}-\mathbf{b} \right\|_2^2 +\lambda | \mathbf{z}| } \;\;\;\;\; {\rm s. } {\rm t.} \; \mathbf{Fx-z=0} .\nonumber 
\end{equation}
By introducing an auxiliary variable $\mathbf{u} $ and starting with initial guesses $\mathbf{x}^0$, $\mathbf{u}^0=\mathbf{z}^0=\mathbf{Fx}^0$, the following iterations are performed to update $\mathbf{x}$ and $\mathbf{z}$:
\begin{align}
\label{eq:dmm_update}
\mathbf{x}^{k+1} &=\left( \mathbf{A}^T\mathbf{A}+s\mathbf{F}^T \mathbf{F}\right)^{-1} (   \mathbf{A}^T\mathbf{b}+   s\mathbf{F}^T(\mathbf{z}^k-\mathbf{u}^k) )\nonumber \\
\mathbf  {z}^{k+1} &={\rm shrink}_{\lambda/s}(\mathbf{Fx}^{k+1}+\mathbf{z}^{k}+\mathbf{u}^k)\\
\mathbf  {u}^{k+1} &=\mathbf{u}^k+\mathbf{Fx}^{k+1}-\mathbf{z}^{k+1}.\nonumber
\end{align}
Here $s>0$ is an augmented lagrangian parameter, and the soft thresholding operator is defined as
\begin{equation}
{\rm shrink}_{\lambda/s}(a)= \left\{ \begin{array}{l}
 a-\lambda/s,\; {\rm{if}}\; a> \lambda/s \\ 
 0 ,\;\;\;\;\; \;\;\;\;\;\;{\rm{if}}\;|a|< \lambda/s \\ 
 a+\lambda/s, \; {\rm{if}}\; a<- \lambda/s. 
 \end{array} \right.\nonumber
\end{equation}

The pseudo codes for solving (\ref{eq:alm_k}) are given in Alg.~\ref{alg:admm}.
\begin{algorithm}[t]
\caption{ADMM for Solving (\ref{eq:alm_k}).}
\label{alg:admm}
\begin{algorithmic}[1]
\STATE {Initialize:  form $\mathbf{A,F}$ and $\mathbf{b}$ according to Appendix~\ref{app:glasso}, specify initial guess $\mathbf{x}^0$, $\mathbf{u}^0$ and $\mathbf{z}^0$;}
\STATE {\textbf{for} $j=0,1,\cdots$ \textbf{do}}
   \STATE {\hspace{5pt} compute $\mathbf{x}^{j+1}$, $\mathbf{z}^{j+1}$ and $\mathbf{u}^{j+1}$ according to (\ref{eq:dmm_update})};
      \STATE {\hspace{5pt} \textbf{break} if $\|\mathbf{Fx}^{j+1}-\mathbf{z}^{j+1}\|<\epsilon_1$ \& $\|\mathbf{F}^T(\mathbf{z}^{j+1}-\mathbf{z}^{j})\|<\epsilon_2$;}
  \STATE {\textbf{end for} } 
\STATE {\textbf{return} $\mathbf{U}^{(k),l+1}={\rm reshape}(\mathbf{x}^{j+1}, [m,r])$ . }
\end{algorithmic}
\end{algorithm}

\subsection{Limitations}
\label{subsec:limitation}
Firstly, the cost function of (\ref{eq:t_recovery}) is non-convex, and it is non-trivial to compute its global minimum with theoretical guarantees. Although researchers and engineers are very often satisfied with a local minimal, the obtained result may not be good enough for some cases. Secondly, in this work the parameters $\lambda$ [the regularization parameter in (\ref{eq:t_recovery})] and $r$ [the tensor rank in \eqref{tensor_lr}] are set based on some heuristic experiences. This treatment is definitely not optimal and does not guarantee high accuracy for all cases. 

\begin{figure}[t]
    \centering
    \includegraphics[width=2.6in]{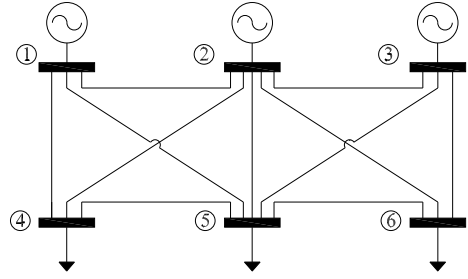}
	\caption{The $6$-bus system.}
     \label{fig:6bus}
\end{figure}

\section{Simulations}
 \label{sec:simulation}
This section reports the simulation results for several test cases from MATPOWER $5.1$~\cite{matpower:2011}. All codes are implemented in MATLAB. 
We find that a $2$nd- or $3$rd-order generalized polynomial-chaos expansion can provide good accuracy for many cases, therefore we set $p=2$ (or $3$) in (\ref{surrogate_i}) and $m=3$ (or $4$) in Equation (\ref{c_tp_inner}) . 
\begin{figure}[t]
    \centering
    \includegraphics[width=3.3in]{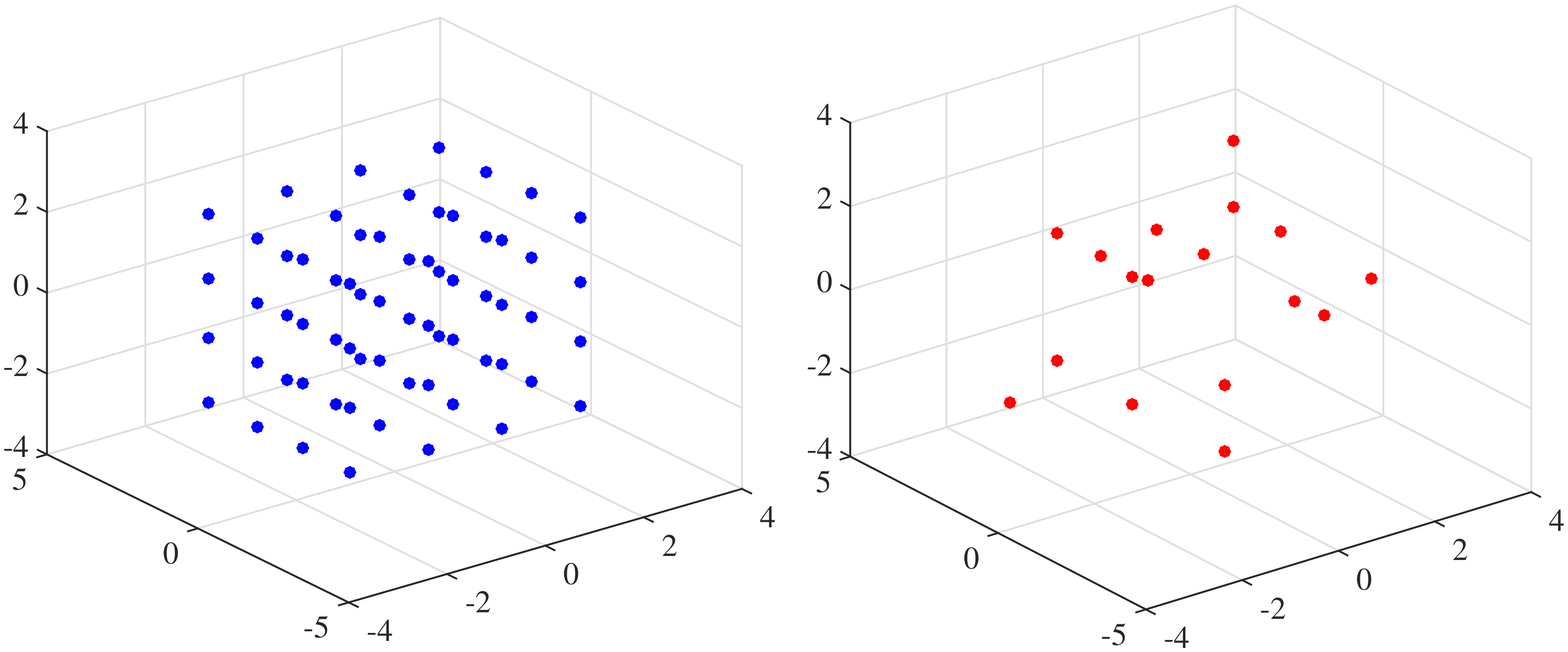}
	\caption{Left: all quadrature samples; right: samples used in tensor recovery.}
     \label{fig:3d_samples}
\end{figure}

\begin{figure*}[t]
    \centering
    \includegraphics[width=5.7in]{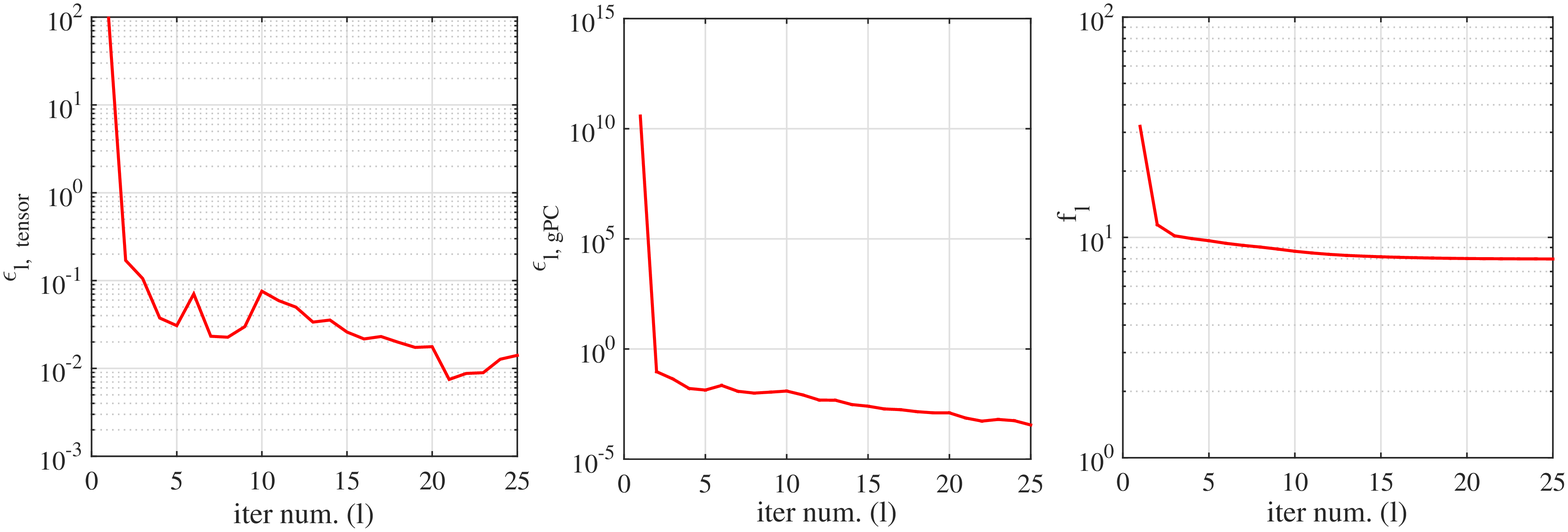}
	\caption{Convergence for the 6-bus example. The left figure shows the relative update of tensor factors (i.e., $\epsilon_{l, \rm tensor}$); the middle figure shows the update of the generalized-polynomial-chaos coefficients (i.e., $\epsilon_{l, \rm gPC}$); the right figure shows the value of the objective function (i.e., $f_l$).}
     \label{fig:6bus_convg}
\end{figure*}
 
\subsection{$6$-Bus Case (with $3$ Random Parameters)}
The \textbf{case6ww} example in MATPOWER 5.1 (c.f. Fig. \ref{fig:6bus}) is used as a demonstrative example. We use $3$ random parameters to describe the uncertain active powers at the load buses $4$ to $6$. We aim to obtain a 3rd-order generalized polynomial-chaos expansion for the real power injected from Bus $2$ to Bus $4$, leading to $20$ basis functions in total. Applying a $4$-point Gauss-quadrature rule to perform numerical integration for each dimension, we generate $64$ quadrature points in total. 

\begin{figure}[t]
    \centering
    \includegraphics[width=3.3in, height=1.7in]{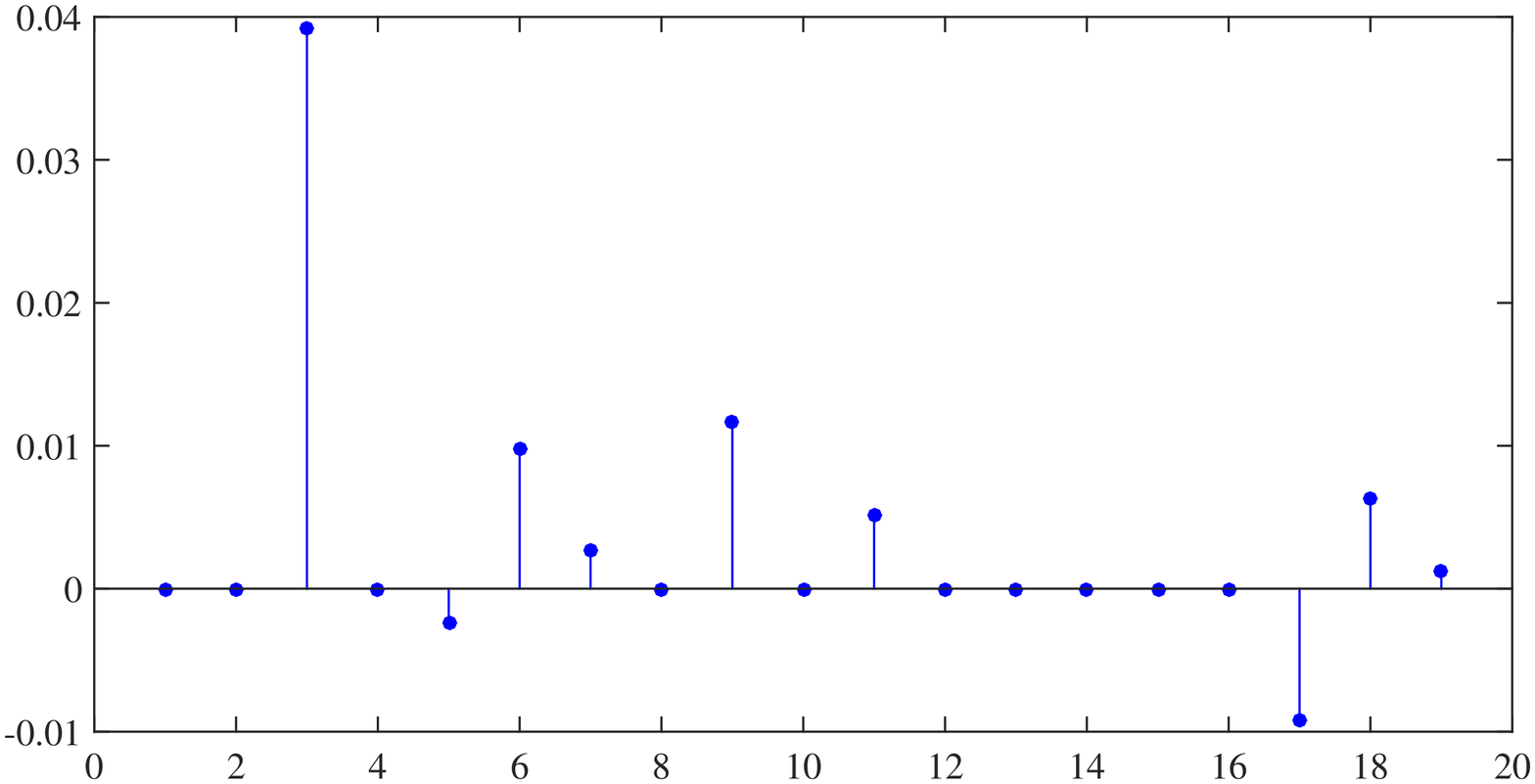}
	\caption{Coefficients of $\{\Psi_{\boldsymbol{\alpha}}(\boldsymbol{\xi})\}_{|\boldsymbol{\alpha}|=1}^3$ for the $6$-bus example.}
     \label{fig:6bus_gPC}
\end{figure}

In order to compute the generalized polynomial-chaos expansion, only $18$ quadrature points (as shown in Fig.~\ref{fig:3d_samples}) are randomly sub-selected. The simulation results at these selected samples are used to perform tensor recovery. For this case, we find that setting the tensor rank $r=3$ and the regularization parameter $\lambda=0.25$ is a good choice. Starting from a randomly generated rank-$3$ tensor, our algorithm converged after $25$ iterations as shown in Fig. ~\ref{fig:6bus_convg}. The obtained low-rank tensor approximation has an estimated prediction error of $0.2\%$. With the obtained tensor approximation, the coefficients for all generalized polynomial-chaos basis functions are easily calculated based on~(\ref{eq:tensor2gPC}). The coefficients for $\boldsymbol{\alpha}=0$ is 31.83, which is the mean value of the output. All other coefficients are plotted in Fig.~\ref{fig:6bus_gPC}, where a sparsity pattern is observed. 

Next we validate our results by Monte Carlo. Here we use $5000$ samples in Monte Carlo simulation and treat its result as a golden reference solution. As shown in Table \ref{table:6bus}, the mean value and standard deviation from our tensor recovery approach is very close to that from Monte Carlo. 

\textbf{Complexity Reduction.} Since we use $18$ samples out of $64$ quadrature points, the reduction ratio for this problem is $3.6$. Note that the number of samples in tensor recovery is less than the number of basis functions (i.e., $20$). 

\begin{table}[t]
    \caption{Moments of the $6$-bus example.}
    \centering
    \label{table:6bus}
    \begin{tabular}{|c |c |c |}
    \hline
   & Tensor Recovery  & Monte Carlo 
              \\  \thickhline
              samples & 18 & 5000 \\ \hline
                  Mean &  31.83  & 31.87  \\ \hline
    stand. dev. & 0.0439 & 0.0448 \\
    \hline
    \end{tabular}
\end{table}


\subsection{$30$-Bus Case (with $24$ Random Parameters)}
Next we consider the \textbf{case30} example in MATPOWER 5.1, with the active powers of $24$ load buses modeled by Gaussian random variables. We apply a $2$nd-order generalized polynomial-chaos expansion for the real power from bus $15$ to bus $23$, requiring totally $325$ basis functions. For each parameter, $3$ quadrature points are used, leading to $3^{24}\approx 2.8\times 10^{11}$ samples in total. Obviously, it is prohibitively expensive to simulate the power system at all quadrature points. 

\begin{figure}[t]
    \centering
    \includegraphics[width=3.3in, height=1.7in]{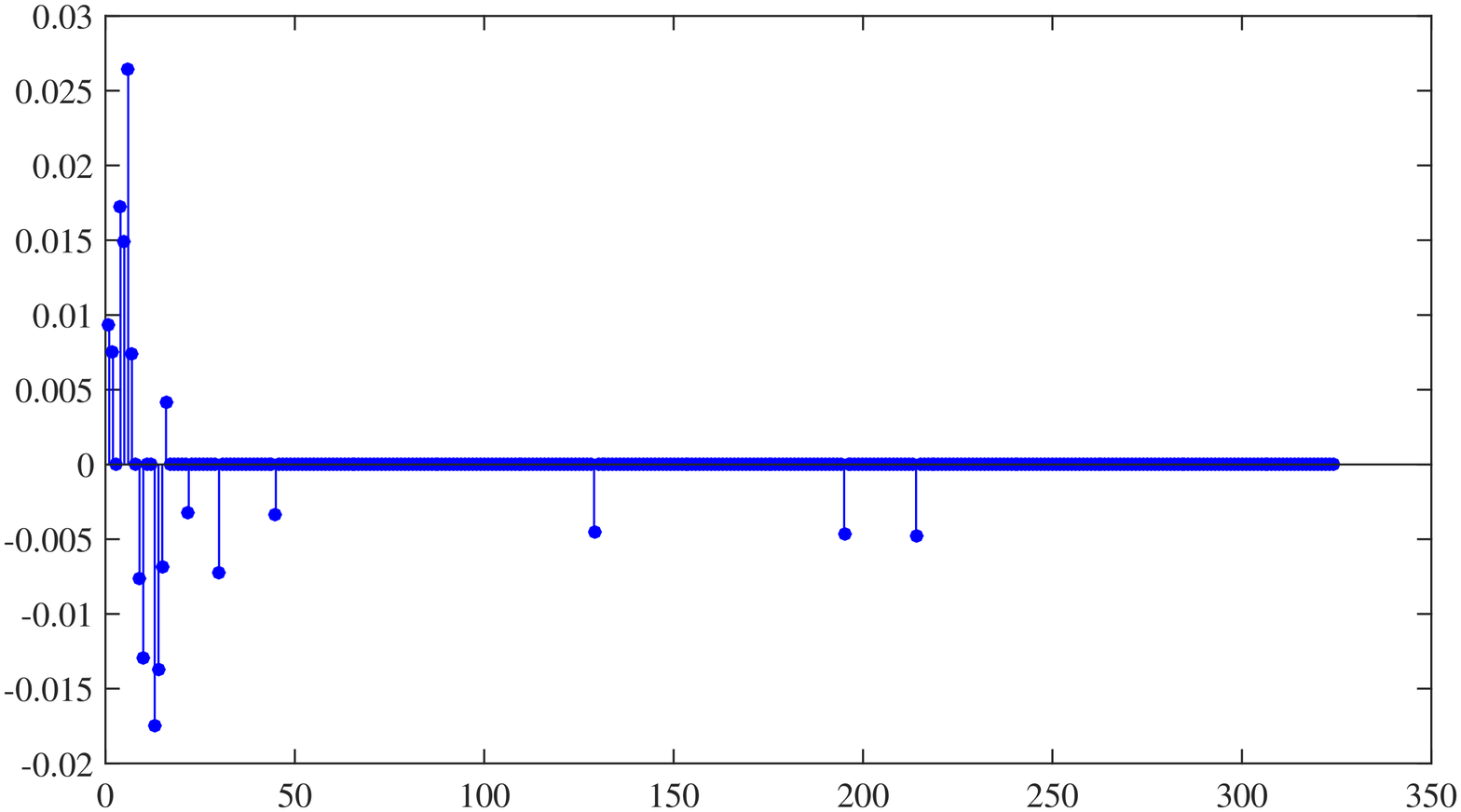}
	\caption{Coefficients of $\{\Psi_{\boldsymbol{\alpha}}(\boldsymbol{\xi})\}_{|\boldsymbol{\alpha}|=1}^2$ for the $30$-bus example.}
     \label{fig:30bus_gPC}
\end{figure}

In our tensor recovery scheme, we randomly pick $280$ quadrature points from the full tensor-rule quadrature samples and approximate $\mathbfcal{G}$ by a rank-$4$ tensor. Setting $\lambda=0.3$ and starting with a random initial guess, our algorithm converges nicely' after $26$ iterations which are similar to Fig.~\ref{fig:6bus_convg}. With $50$ newly sub-sampled quadrature points as the testing samples, the estimated prediction error is $0.55\%$. Although this example has many random parameters, its generalized polynomial-chaos expansion is very sparse, as shown in Fig.~\ref{fig:30bus_gPC}.


\begin{figure}[t]
    \centering
    \includegraphics[width=3.3in]{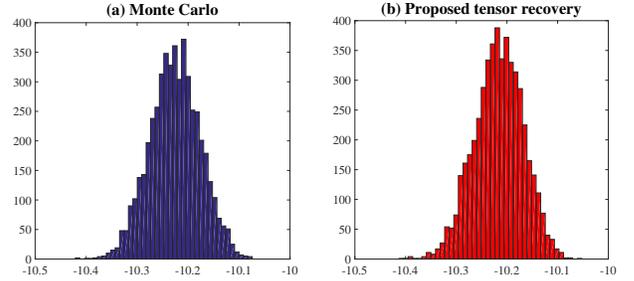}
	\caption{Histograms of the simulated output for the 30-bus example .}
     \label{fig:30bus_hist}
\end{figure}

In order to check the accuracy, we perform Monte Carlo simulation using $5000$ random samples. Table~\ref{table:30bus} compares the mean values and standard deviations from both approaches, and they are very close.  An advantage of generalized polynomial-chaos expansion is that one can easily evaluate the expression with many samples to get a density function or histogram. Such information cannot be easily obtained by a point-estimation method. The histogram from our method is close to that from Monte Carlo (c.f. Fig. ~\ref{fig:30bus_hist}).

\begin{figure*}[t]
    \centering
    \includegraphics[width=5.7in]{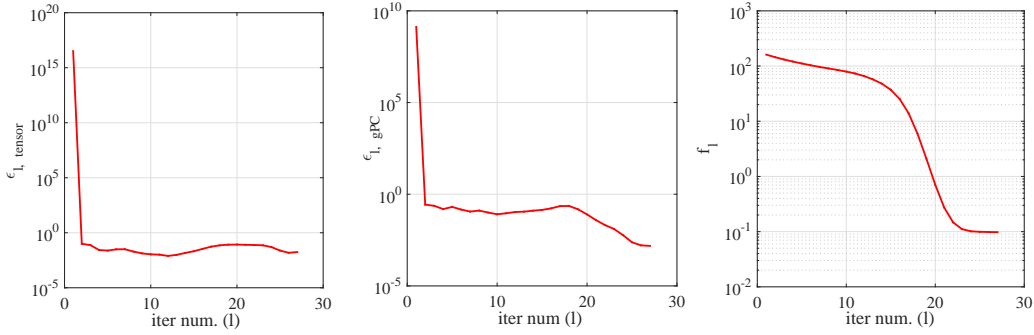}
	\caption{Convergence for the $57$-bus example. The left figure shows the relative update of tensor factors (i.e., $\epsilon_{l, \rm tensor}$); the middle figure shows the update of the generalized-polynomial-chaos coefficients (i.e., $\epsilon_{l, \rm gPC}$); the right figure shows the value of the objective function (i.e., $f_l$).}
     \label{fig:57bus_convg}
\end{figure*}
\begin{figure}[t]
    \centering
    \includegraphics[width=3.3in, height=1.7in]{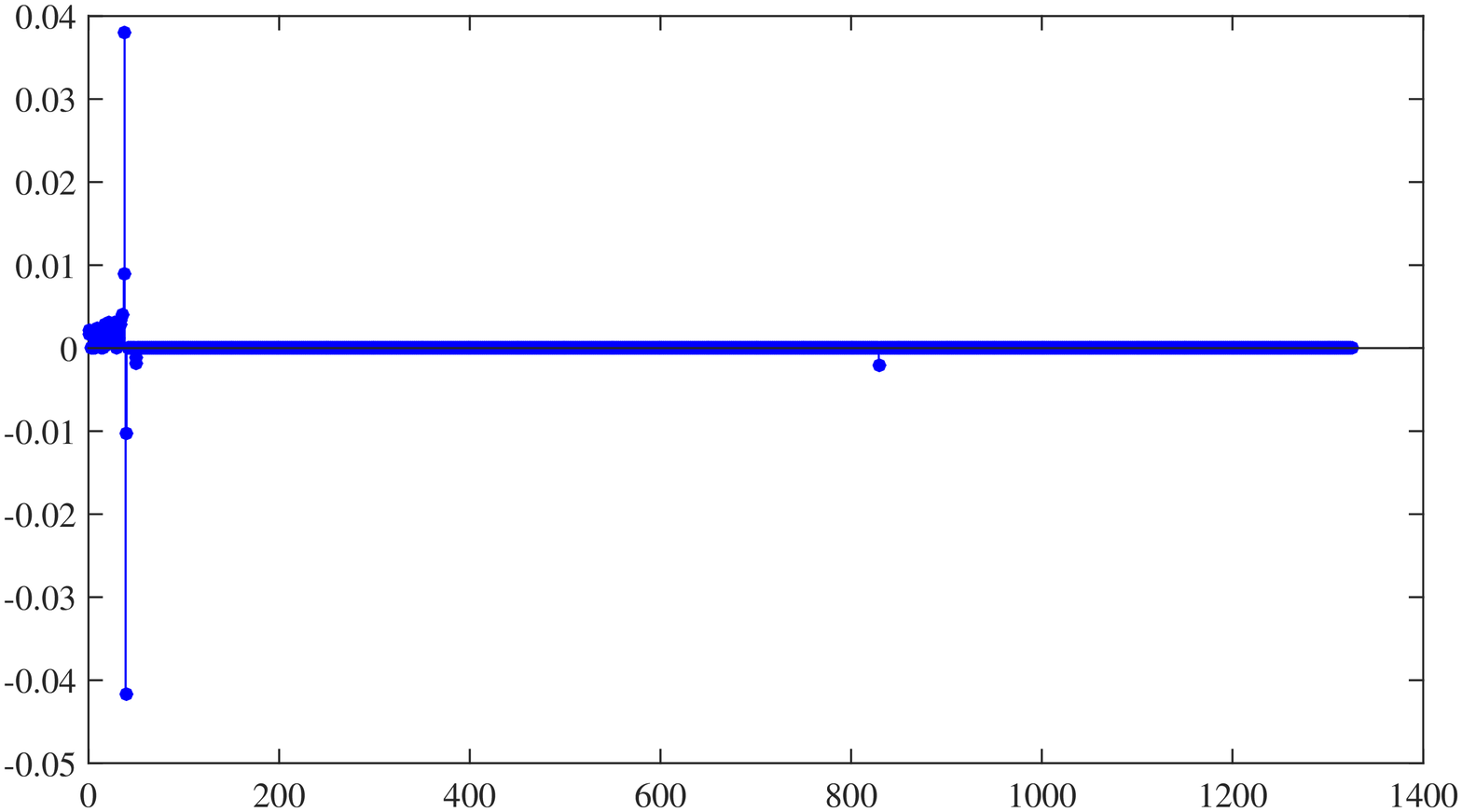}
	\caption{Coefficients of $\{\Psi_{\boldsymbol{\alpha}}(\boldsymbol{\xi})\}_{|\boldsymbol{\alpha}|=1}^2$ for the $57$-bus example.}
     \label{fig:57bus_gPC}
\end{figure}
\begin{table}[t]
    \caption{Moments of the $30$-bus example.}
    \centering
    \label{table:30bus}
    \begin{tabular}{|c |c |c |}
    \hline
   & Tensor Recovery  & Monte Carlo 
              \\  \thickhline
              samples & 280 & 5000 \\ \hline
                  Mean &  -10.23  & -10.22  \\ \hline
    stand. dev. & 0.048 & 0.049 \\
    \hline
    \end{tabular}
\end{table}

\textbf{Complexity Reduction.} Since we use $280$ samples out of $3^{24}$ quadrature points, the reduction ratio for this example is $10^9$. The number of samples in tensor recovery is also smaller than the number of basis functions (i.e., $325$).

\subsection{$57$-Buse Case (with $50$ Random Parameters)}
Finally we consider the \textbf{case57} example in MATPOWER 5.1,  with $50$ Gaussian random variables describing the active powers at load buses. With a $2$nd-order polynomial-chaos expansion, we aim to approximate the real power injected from Bus $19$ to Bus $20$ with $1326$ basis functions. Using $3$ Gauss-quadrature points for each parameter, a tensor-rule quadrature method requires $3^{50}>7\times 10^{23}$ samples in total. It is impossible to store the samples on a personal computer, let alone simulating the power flow equation at all samples.

Our tensor recovery scheme randomly sub-selects $800$ samples to perform power flow simulations. Starting with a random initial guess, we approximate the full tensor $\mathbfcal{G}$  by a rank-$5$ tensor, with an estimated prediction error of $1\%$. Fig. ~\ref{fig:57bus_convg} shows the convergence of our solver.  Fig.~\ref{fig:57bus_gPC} plots the coefficients for all non-constant basis functions. Clearly, the result is extremely sparse for this high-dimensional example.

\begin{figure}[t]
    \centering
    \includegraphics[width=3.3in, height=1.7in]{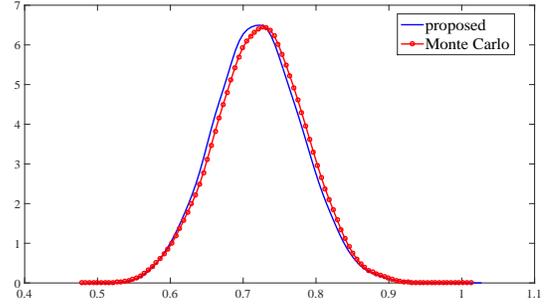}
	\caption{Computed probability density functions for the $57$-bus example.}
     \label{fig:57bus_pdf}
\end{figure}

In order to get a full picture about the statistical behavior of the output, we evaluate the computed generalized polynomial-chaos expansion with $5000$ random samples and plot its probability density function. As shown in Fig.~\ref{fig:57bus_pdf}, the result is close to that from Monte Carlo simulation on the original power flow equations. The mean values and standard deviations from both approaches are very close (c.f.  Table~\ref{table:57bus}).

\textbf{Complexity Reduction.} Since we use $800$ samples out of $3^{50}$ quadrature points, the reduction ratio for this example is about $9\times 10^{20}$. The number of samples in tensor recovery is again smaller than the number of basis functions (i.e., $1326$).
\begin{table}[t]
    \caption{Moments of the 57-bus example.}
    \centering
    \label{table:57bus}
    \begin{tabular}{|c |c |c |}
    \hline
   & Tensor Recovery  & Monte Carlo 
              \\  \thickhline
              samples & 800 & 5000 \\ \hline
                  Mean &  0.721 & 0.724  \\ \hline
    stand. dev. & 0.0596 & 0.0609 \\
    \hline
    \end{tabular}
\end{table}

\section{Conclusions and Future Work}\label{sec:futurework}
This paper has presented a probabilistic power flow simulation algorithm based on tensors and stochastic collocation. In order to break the curse of dimensionality, we have developed a high-dimensional method that exploits the low-rank and sparse property of tensors. This tensor framework has completed the huge-data-size simulation task with an extremely small-size simulation data set. We have further developed a numerical solver for the tensor recovery problem and tested it on three power flow benchmarks. This algorithm has successfully generated high-dimensional and sparse generalized polynomial-chaos expansions for the solutions. Good accuracy (measured by prediction errors and comparison against Monte Carlo)  as well as significant computational cost reduction (with up to $9\times 10^{20}$ times) have been observed in this work.

To the best of our knowledge, this is the first work studying stochastic power systems from a tensor perspective. While several limitations exist in the current work, the authors believe that many problems are worth investigation in this direction. Firstly, it is worth investing global non-convex optimization to solve (\ref{eq:t_recovery}) or trying to convexify the formulation. Secondly, some future work about the optimal choice of $\lambda$ and $r$ may help improve the robustness of our framework. Finally, it is worth developing better sampling schemes to improve the performance of tensor recovery. 


One particularly attractive application of this technique is the construction of probabilistic static equivalents of a sub-networks. These equivalents can be used both for distribution grid models with high penetration of intermittent renewables and for probabilistic modeling of random disturbances in neighbor areas in multi-area power systems. 

A second natural application is the stochastic contingency analysis over a range of operating conditions. Currently deterministic contingency analysis is considered in some basic cases with given operating condition. However, as the systems change and are subject to uncertainties, one need to take multiple cases or scenarios into account. The fast averaging technique developed in this paper can effectively alleviate the heavy computation in such situations.


\appendices
\section{Orthonormal Polynomials}
\label{subsec:uni_gPC}
Consider a single random parameter $\xi_k \in \mathbb{R}$ with a probability density function $\rho_k(\xi_k)$, one can construct a set of polynomial functions subject to the orthonormal condition:
\begin{equation}
\label{uni_gPC}
\int\limits_{\mathbb{R}}  {\phi_{k,\alpha} ( {\xi_k } )\phi_{k,\beta} ( {\xi_k } ){\rho_k}( {\xi_k } )d\xi_k }=\delta_{\alpha,\beta} \nonumber
\end{equation}
where  $\delta_{\alpha,\beta}$ is a Delta function, integer $\alpha$ is the highest degree of $\phi_{k,\alpha} ( {\xi_k } )$. Such polynomials can be constructed as follows~\cite{Walter:1982}. First, one constructs orthogonal polynomials $\{\pi_{k, \alpha}(\xi_k) \}_{\alpha=0}^p$ with an leading coefficient 1 recursively
\begin{equation}
\label{recurrence}
 \pi _{k,\alpha + 1} (\xi_k) = \left( {\xi_k - \gamma _{\alpha} } \right)\pi_{k,\alpha} (\xi_k) - \kappa _{\alpha} \pi _{k, \alpha - 1} (\xi_k)\nonumber
\end{equation}
 for $\alpha = 0,1, \cdots p-1$, with initial conditions $\pi_{k, - 1} (\xi_k) = 0$, $\pi_{k,0} (\xi_k) = 1$  and $\kappa_0=1$. For $\alpha\geq 0$, the recurrence parameters are defined as
\begin{equation}
\label{int_cal}
\begin{array}{l}
 \displaystyle{\gamma _{\alpha}  = \frac{  {\mathbb{E}\left(\xi_k\pi _{k,\alpha}^2 (\xi_k)\right)}  }{{  {\mathbb{E}\left(\pi _{k,\alpha}^2 (\xi_k)\right)} }}}, \;\displaystyle{\kappa _{\alpha+1}  = \frac{   {\mathbb{E}\left(\xi_k\pi_{k,\alpha+1}^2 (\xi_k)\right)}   }{   {\mathbb{E}\left(\xi_k\pi _{k,\alpha}^2 (\xi_k)\right)} }}. 
 \end{array}
\end{equation}
Here $\mathbb{E}$ denotes the operator that calculates expectation. Second, one can obtain $\{\phi_{k, \alpha}(\xi_k) \}_{\alpha=0}^p$ by normalization:
\begin{equation}
\phi_{k,\alpha}(\xi_k) = \frac{{\pi_{k,\alpha}(\xi_k)}}{{\sqrt {\kappa _0 \kappa _1  \cdots \kappa _{\alpha} } }}, \; {\rm for}\; \alpha=0,1,\cdots,p. \nonumber
\end{equation}

\section{Gauss Quadrature Rule~\cite{Golub:1969}}
\label{app:gauss_quad}
Given $\xi_k \in \mathbb{R}$ with a density function $\rho_k(\xi_k)$ and a smooth function $q(\xi_k)$,  Gauss quadrature evaluates the integral
\begin{equation}
\label{stoInt}
\int\limits_{\mathbb{R} } {q( {\xi _k } )\rho_k ( {\xi _k } )d\xi _k }  \approx \sum\limits_{i_k = 1}^{m} {q( {\xi _k^{i_k} } )} w_k^{i_k} \nonumber
\end{equation}
with an error decreasing exponentially as $m$ increases. An exact result is obtained if $q(\xi_k)$ is a polynomial function of degree $\leq 2 m-1$. One can obtain $\{(\xi_k^{i_k}, w_k^{i_k})\}_{i_k=1}^{p+1}$ by reusing the recurrence parameters in (\ref{int_cal}) to form a symmetric tridiagonal matrix $\mathbf{J} \in \mathbb{R}^{(p+1)\times (p+1)}$: 
\begin{equation}
\label{eq:jmatrix}
\mathbf{J}\left( {i,j} \right) = \left\{ \begin{array}{l}
 \gamma_{i - 1} ,\;{\rm{if}}\;i = j \\ 
 \sqrt {\kappa _{i} } ,\;{\rm{if}}\; i = j+ 1 \\ 
 \sqrt {\kappa _{j} } ,\;{\rm{if}}\; i = j - 1 \\ 
 0,\;{\rm{otherwise}} \\ 
 \end{array} \right.\;{\rm{for}}\;1 \le i,j \le p + 1. \nonumber
\end{equation}
Let $\mathbf{J} = \mathbf{Q}\Sigma \mathbf{Q}^T$ be an eigenvalue decomposition and $\mathbf{Q}$ a unitary matrix, then $\xi_k^{i_k}=\Sigma(i_k,i_k)$ and $w_k^{i_k}=\left(\mathbf{Q}(1,i_k)\right)^2$.

\section{Error Control in Alg.~\ref{alg:alm}} 
\label{app:errDef}
With tensor factors $\{ \mathbf{U}^{(1), k}\}_{k=1}^d$ obtained after $l$ iterations of the outer loops of Alg.~\ref{alg:alm}, we define
\small
\begin{align}
f_l & :=f\left( \mathbf{U}^{(1), l}, \cdots,   \mathbf{U}^{(d), l} \right) &   [{\rm updated \; cost \; func. \; of}\; (\ref{eq:t_recovery})] \nonumber \\
\hat{\mathbfcal{G}} _l & :=\mathbb{T}  \left( \mathbf{U}^{(1),l}, \cdots, \mathbf{U}^{(d),l} \right) & (\rm approximated \; tensor) \nonumber\\
c_{\boldsymbol{\alpha}}^l&:=\left\langle \hat{\mathbfcal{G}}_l, \mathbfcal{W}_{\boldsymbol{\alpha}} \right\rangle & [{\rm updated \; coefficient\; for} \; \Psi_{\boldsymbol{\alpha}} (\boldsymbol{\alpha})]\nonumber
\end{align} \normalsize
and let $\mathbf{c}^l =[\cdots,  c_{\boldsymbol{\alpha}}^l, \cdots ]\in \mathbb{R}^K$ for all $|\boldsymbol{\alpha}|\leq p$.  Then, we define the following quantities for error control:
 \begin{itemize}
 \item Relative update of the tensor factors:
 \begin{align} 
\epsilon_{l, \rm tensor} & =\sqrt{{\sum \limits_{k=1}^d{\| \mathbf{U}^{(k),l} -\mathbf{U}^{(k),l-1} \|_F^2} }/ { \sum \limits_{k=1}^d{\| \mathbf{U}^{(k),l-1}} \|_F^2}  } .\nonumber
\end{align}

\item Relative update of $\mathbf{c}=[\cdots, c_{\boldsymbol{\alpha}},\cdots]$
 \begin{align} 
\epsilon_{l, \rm gPC}  = { \|  \mathbf{c}^l-  \mathbf{c}^{l-1} \|   }  /  {\| \mathbf{c}^{l-1}\|} .  \nonumber
\end{align} 

\item Relative update of the cost function:
 \begin{align} 
\epsilon_{l,\rm cost} &= { |  f_l-   f_{l-1} | }  /  {|f_{l-1}|}. \nonumber
\end{align} 

 \end{itemize}
The solution $\{ \mathbf{U}^{(k), l}\}_{k=1}^d$ is regarded as a local minimal if $\epsilon_{l, \rm tensor} $, $\epsilon_{l, \rm gPC} $ and $\epsilon_{l, \rm cost} $ are small enough.

\section{Assembling The Matrices and Vector in (\ref{eq:glasso})}
\label{app:glasso}
Consider the tensor factors $\mathbf{U}^{(1),l+1}$, $\cdots$, $\mathbf{U}^{(k-1),l+1}$, $\mathbf{X}$, $\mathbf{U}^{(k+1),l}$, $\cdots$, $\mathbf{U}^{(d),l}$ in (\ref{eq:alm_k}). We denote the $(i,j)$ element of $\mathbf{U}^{(k'),l}$ (or $\mathbf{X}$)  by $u_{i,j}^{(k'),l}$ (or $x_{i,j}$), and its $j$-th column by  $\mathbf{u}_{j}^{(k'),l}$ (or $\underline{\mathbf{x}}_j$). Then, the cost function in (\ref{eq:alm_k}) is
\small
\begin{align}
&f \left( \cdots, \mathbf{U}^{(k-1),l+1}, \mathbf{X}, \mathbf{U}^{(k+1),l},\cdots \right) \nonumber\\
=& \frac{1}{2}\sum \limits_{\mathbf{i}\in \Omega} { \left(\sum \limits_{j=1}^r {x_{i_k,j} }\mu_{\mathbf{i},j}-\mathbfcal{G} (\mathbf{i}) \right)^2  } +\lambda \sum \limits_{|\boldsymbol{\alpha}|\leq p} {\left|   \sum \limits_{j=1}^r {   \nu_{\boldsymbol{\alpha}, j} 	\langle \underline{\mathbf{x}}_j, \mathbf{w}_{\alpha _k}^{(k)} \rangle} \right|} \nonumber
\end{align} \normalsize
where the scalars $\mu_{\mathbf{i},j}$ and $ \nu_{\boldsymbol{\alpha}, j}  $ are computed as follows:
\begin{align}
\mu_{\mathbf{i},j} &= \prod \limits_{k'=1} ^{k-1}{  u_{i_{k'},j}^{(k'),l+1} }\prod \limits_{k'=k+1} ^{d}{  u_{i_{k'},j}^{(k'),l} }, \nonumber \\
 \nu_{\boldsymbol{\alpha}, j} &= \prod \limits_{k'=1}^{k-1} { \langle \mathbf{u}_{j}^{(k'),l+1}, \mathbf{w}_{\alpha _{k'}}^{(k')} \rangle}   \prod \limits_{k'=k+1}^{d} { \langle \mathbf{u}_{j}^{(k'),l}, \mathbf{w}_{\alpha _{k'}}^{(k')} \rangle}. \nonumber
\end{align}
 
  Since each row (or element) of $\mathbf{A}$ (or $\mathbf{b}$) corresponds to an index $\mathbf{i}\in \Omega$, and each row of $\mathbf{F}$ corresponds to a basis function $\Psi_{\boldsymbol{\alpha}} (\boldsymbol {\xi})$, in this appendix we use $\mathbf{i}$ as the row index (or element index) of $\mathbf{A}$ (or $\mathbf{b}$) and $\boldsymbol{\alpha}$ as the row index of $\mathbf{F}$. Now we specify the elements of $\mathbf{A}$, $\mathbf{b}$ and $\mathbf{F}$ of  (\ref{eq:glasso}).
  \begin{itemize}
  \item For every $\mathbf{i} \in \Omega$, $\mathbf{b}(\mathbf{i})=\mathbfcal{G} (\mathbf{i})$.
  
  \item Since $x_{i_k,j}$ is the $(j-1)m+ i_k$-th element of $\mathbf{x}=\boldsymbol{vec}  (\mathbf{X})$, for every $\mathbf{i} \in \Omega$ we have
  \begin{equation}
\mathbf{A}(\mathbf{i},(j-1)m+i_k)= \left\{ \begin{array}{l}
 \mu_{\mathbf{i},j},\; {\rm{for}}\; j=1,\cdots,r \\ 
 0 ,\;\;\;\;{\rm otherwise}. 
 \end{array} \right.\nonumber
\end{equation}
  
  \item Since $\underline{\mathbf{x}}_j$ includes the elements of $\mathbf{x}$ ranging from index $(j-1)m+1$ to $jm$, given an index vector $\boldsymbol{\alpha}$ the corresponding row of $\mathbf{F}$ can be specified as 
  \begin{equation}
  \mathbf{F}(\boldsymbol{\alpha},jm-m+i_k)= \nu_{\boldsymbol{\alpha},j}\mathbf{w}_{\alpha_k}^{(k)}(i_k)=\nu_{\boldsymbol{\alpha},j} \phi_{k,\alpha_k}(\xi_k^{i_k})w_k^{i_k}    \nonumber
  \end{equation}
 for all integers $j\in [1,r]$ and $i_k\in [1,m]$.
 
  \end{itemize}

\bibliographystyle{IEEEtran}
\bibliography{main}

\end{document}